\documentclass[%
 reprint,
 prl,
 amsmath,
 amssymb,
 aps,
 floatfix,
]{revtex4-1}

\usepackage{graphicx}
\usepackage{dcolumn}
\usepackage{bm}
\usepackage{hyperref}
\usepackage[mathlines]{lineno}
\usepackage{setspace}
\usepackage{physics}
\usepackage{mathtools}
\usepackage{xcolor}

\begin{document}
\title{Photonic modal circulator using temporal refractive-index modulation with spatial inversion symmetry}
\author{Jiahui Wang}
\author{Jason F. Herrmann}
\author{Jeremy D. Witmer}
\author{Amir H. Safavi-Naeini}
\affiliation{Department of Applied Physics, Stanford University, Stanford, CA 94305 USA}
\author{Shanhui Fan}%
\email{shanhui@stanford.edu}
\affiliation{Department of Electrical Engineering, Stanford University, Stanford, CA 94305 USA}

\begin{abstract}
It has been demonstrated that dynamic refractive index modulation, which breaks time-reversal symmetry, can be used to create on-chip non-reciprocal photonic devices. In order to achieve amplitude non-reciprocity, all such devices moreover require modulations that break spatial symmetries, which adds complexity in implementations. Here we introduce a modal circulator, which achieves amplitude non-reciprocity through a circulation motion among three modes. 
We show that such a circulator can be achieved in a dynamically-modulated structure that preserves mirror symmetry, and as a result can be implemented using only a single standing-wave modulator, which significantly simplifies the implementation of dynamically-modulated non-reciprocal device. 
We also prove that in terms of the number of modes involved in the transport process, the modal circulator represents the minimum configuration in which complete amplitude non-reciprocity can be achieved while preserving spatial symmetry. 
\end{abstract}
                              
\maketitle


\textit{Introduction.} The explorations of non-reciprocal photonic structures~\cite{potton2004reciprocity,fanSci2012comment,caloz2018electromagnetic,williamson2020integrated,jalas2013NatPhotComment} have been of fundamental importance since they offer unique properties, such as optical isolations~\cite{FangPRL2012, Lira2012PRL, YuNatPho2009, tzuang2014natphoton, kittlaus2018phonon, bi2011NatPhoMO}, and robust transport~\cite{fang2012realizing, hafezi2013imaging, raghu2008Haldane, Wang2009Natureobservation} through disordered systems without the need of symmetry protection, that cannot be achieved in reciprocal systems. 
Among various paths for creating non-reciprocal photonic structures, the use of dynamically-modulated non-magnetic systems~\cite{FangPRL2012, peterson2019strongPRL, estep2014NatPhot, Lira2012PRL, doerr2011tandem, YuNatPho2009, sounas2014angular, tzuang2014natphoton, cardea2020arbitrarily}, where the refractive index of the system is modulated as a function of time and space, has been of significant recent interests since it offers a route to create non-reciprocal physics using standard optical materials such as silicon~\cite{reed2010silicon}.

To achieve non-reciprocity through dynamic modulation, both the space and time dependency of the modulation needs to be carefully considered~\cite{FangPRL2012, peterson2019strongPRL, estep2014NatPhot, Lira2012PRL, doerr2011tandem, YuNatPho2009, sounas2014angular, tzuang2014natphoton}. 
Certainly, the modulations must have the appropriate temporal waveforms to break reciprocity.
In addition, all dynamically-modulated on-chip structures considered so far have used a spatial dependency of the modulation that breaks spatial inversion symmetry. 
For example, for non-reciprocal structures based on traveling wave modulators~\cite{YuNatPho2009, kittlaus2018phonon}, the directionality of the traveling wave breaks spatial inversion symmetry.
Similarly, in the optical isolators based on the photonic Aharonov-Bohm effect~\cite{FangPRL2012, tzuang2014natphoton}, the spatial symmetry is broken with the use of two standing-wave modulators with different modulation phases.

In this Letter, we provide a discussion of the requirement on spatial symmetries in dynamically-modulated non-reciprocal systems.
We show that breaking spatial inversion symmetry is indeed required in all systems considered previously~\cite{FangPRL2012, Lira2012PRL, YuNatPho2009,tzuang2014natphoton} to achieve amplitude non-reciprocity, since only two modes are involved in the transport process.
On the other hand, in systems where three  modes are involved in the transport process, a non-reciprocal amplitude response is in fact possible even when the modulated system preserves inversion symmetry. 
As a demonstration of this theoretical understanding, we introduce a non-reciprocal device involving only a single standing-wave modulator, in a structure that preserves mirror symmetry.
This design represents a significant simplification for achieving on-chip non-reciprocal devices based on dynamic modulations.

\begin{figure}[htp!]
    \centering
    \includegraphics[width=\linewidth]{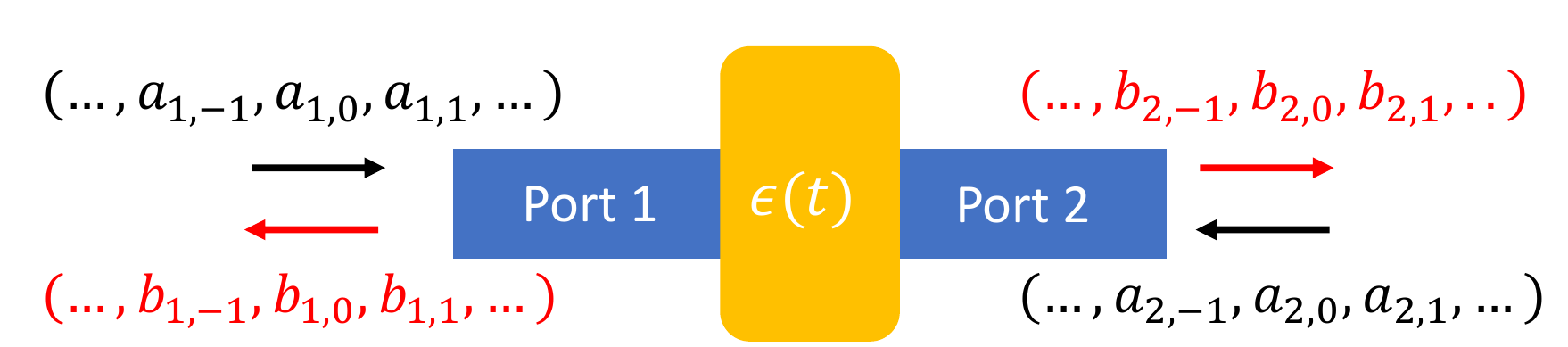}
    \caption{Dynamically-modulated two-port system. The fields at each port consist of multiple frequency sidebands.}
    \label{fig:ports}
\end{figure}

\textit{Scattering Matrix.} We start with a discussion of the implications of inversion symmetry in the construction of non-reciprocal devices based on dynamic modulations.
Suppose the two-port system is harmonically modulated at frequencies that are integer multiples of $\Omega$. In response to incident light at a frequency $\omega$, the steady state consists of multiple sidebands, with the $n$th sideband at the frequency $\omega+n\Omega$, where $n$ is an integer.
At the $i$th port, we denote the incoming and outgoing amplitudes in the $n$th sideband as $a_{i,n}$ and $b_{i,n}$ respectively, with the normalization chosen such that $\abs{a_{i,n}}^2$ and $\abs{b_{i,n}}^2$ correspond to the photon number flux~\cite{FangPRL2012, YuNatPho2009}, as shown in Fig.~\ref{fig:ports}. 
The scattering matrix of the dynamically-modulated two-port system~\cite{williamson2020integrated,tymchenko2017composite} is then
\begin{equation}
    \label{eq:S}
    \begin{pmatrix}
    \mathbf{b}_1\\
    \mathbf{b}_2
    \end{pmatrix} = \mathbf{S}\begin{pmatrix}
    \mathbf{a}_1\\
    \mathbf{a}_2
    \end{pmatrix}, 
\end{equation}
where $\mathbf{a}_{i} = [\dots, a_{i,-1}, a_{i,0}, a_{i, +1}, \dots]^T$ and $\mathbf{b}_{i} = [\dots, b_{i,-1}, b_{i,0}, b_{i, +1}, \dots]^T$.
In what follows, we consider a system with no backscattering, and hence
\begin{equation}
\label{eq:S_TbTf}
    \mathbf{S} = \begin{pmatrix}
    \mathbf{0} & \mathbf{T}_b\\
    \mathbf{T}_f & \mathbf{0}\\
    \end{pmatrix},
\end{equation}
where $\mathbf{T}_f$ and $\mathbf{T}_b$ are the transfer matrices for the forward and backward directions, respectively. 

We define the breaking of the amplitude reciprocity when
\begin{equation}
\label{eq:nonrep_S}
    \abs{\mathbf{S}}_\odot \neq \abs{\mathbf{S}^{T}}_\odot,
\end{equation}
where $\abs{\cdot}_\odot$ represents the absolute-value function operating elementwise on the matrix.
From Eqs.~\eqref{eq:S_TbTf} and \eqref{eq:nonrep_S}, the amplitude non-reciprocity in the system requires that
\begin{equation}
    \label{eq:TfTb}
    \abs{\mathbf{T}_f}_\odot \neq \abs{\mathbf{T}_b^T}_\odot.
\end{equation}
On the other hand, for a structure with either inversion or mirror symmetry that maps one port to the other, we have
\begin{equation}
    \label{eq:TfTbT}
    \mathbf{T}_f = \mathbf{T}_b \equiv \mathbf{T}.
\end{equation}
Thus, to achieve amplitude non-reciprocity, we must have
\begin{equation}
\label{eq:Trequire}
    \abs{\mathbf{T}}_\odot \neq \abs{\mathbf{T}^T}_\odot.
\end{equation}

For lossless dynamically-modulated photonic structures, the photon number flux is conserved~\cite{FangPRL2012, YuNatPho2009}. Thus, the transfer matrix is unitary~\cite{williamson2020integrated}:
\begin{equation}
\label{eq:Tconserve}
    \mathbf{T} \mathbf{T}^\dagger = I.
\end{equation}
In a lossless two-mode system, Eq.~\eqref{eq:Tconserve} implies $\abs{t_{12}} = \abs{t_{21}}$. 
Thus, $\abs{\mathbf{T}}_\odot$ is always symmetric in a two-mode system with inversion or mirror symmetry.
This theoretical result is consistent with Ref.~\cite{FangPRL2012, Lira2012PRL, YuNatPho2009, tzuang2014natphoton}, all of which considered systems with two modes, and utilized modulation schemes that break inversion and mirror symmetry.
On the other hand, the derivation above also indicates that it is in fact possible to achieve amplitude non-reciprocity with three modes in each port, using systems with inversion or mirror symmetry, which provides simpler modulation schemes.
In what follows, we will provide several examples to illustrate this possibility.

\begin{figure}
    \centering
    \includegraphics[width=\linewidth]{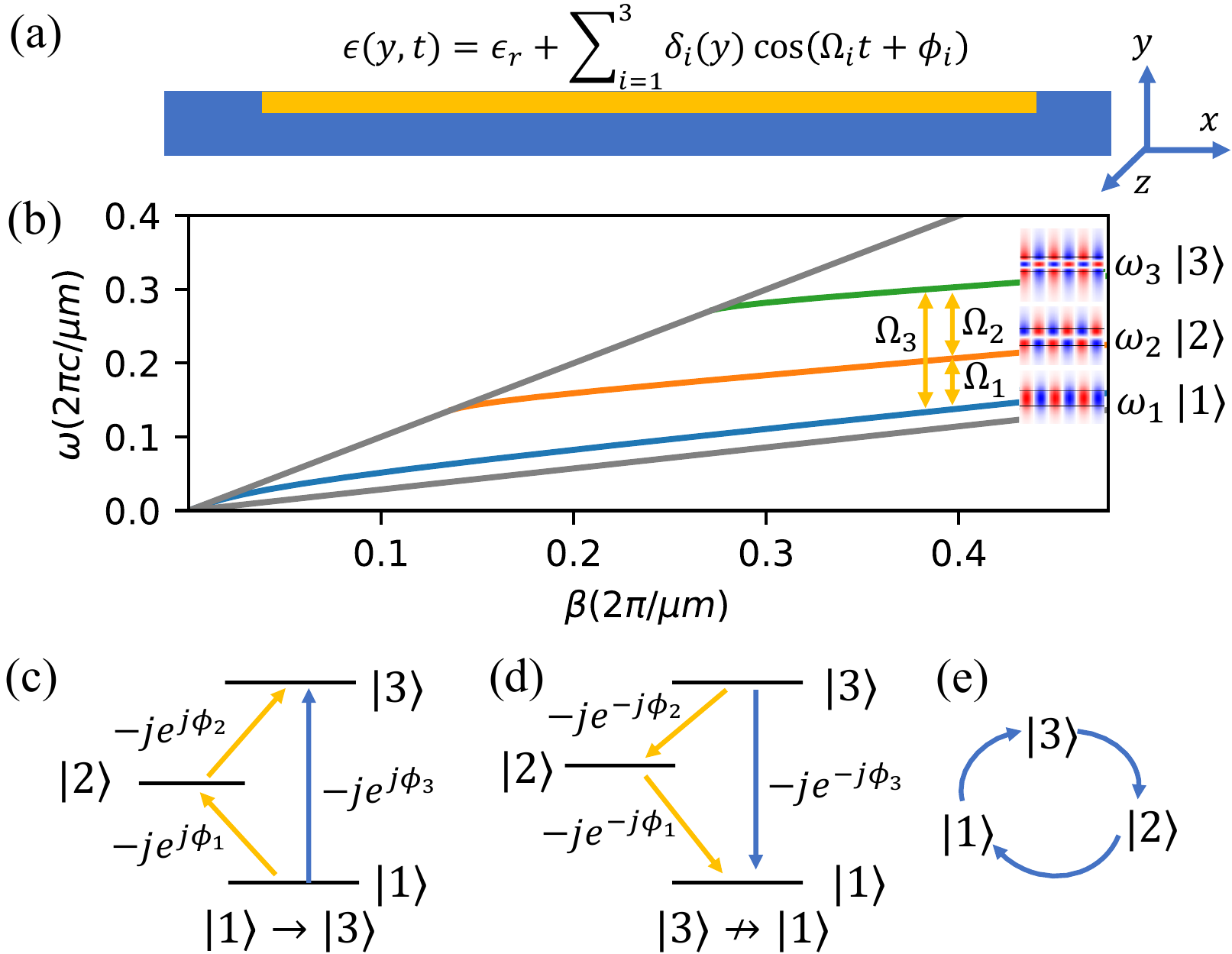}
    \caption{(a) A slab waveguide design. 
    Blue region represents the static waveguide surrounded by air. 
    The modulation is applied to 1/3 of the waveguide shown as the yellow region. 
    (b) Bandstructure of the slab waveguide. The waveguide supports three modes. The modulation frequencies $\Omega_1$, $\Omega_2$, and $\Omega_3$ are chosen to couple $\ket{1}\leftrightarrow\ket{2}$, $\ket{2}\leftrightarrow\ket{3}$, and $\ket{1}\leftrightarrow\ket{3}$, respectively. 
    (c) and (d) show the phase factors associated with transitions between mode $\ket{1}$ and mode $\ket{3}$.
    By choosing modulation phases such that $(\phi_1+\phi_2)-\phi_3=\pi/2$: (c) For the upconversion from mode $\ket{1}\to \ket{3}$, the yellow and the blue pathways constructively interfere.
    (d) For the downconversion from mode $\ket{3}\to \ket{1}$, the two pathways destructively interfere. 
    The transition between mode $\ket{1}$ and $\ket{3}$ is non-reciprocal. 
    (e) Unidirectional circulation among the three modes.}
    \label{fig:waveguide_config_level}
\end{figure}

\textit{Waveguide implementation.} As a first illustration of the three-mode system as indicated above from the scattering matrix analysis, we consider a slab waveguide that supports three TE ($E_z$) modes $\ket{1}$, $\ket{2}$, and $\ket{3}$.
The bandstructure of the waveguide is shown in Fig.~\ref{fig:waveguide_config_level}(b).
The three different modes have three different frequencies $\omega_1$, $\omega_2$, and $\omega_3$ at the same propagation constant $\beta$.
The waveguide is dynamically modulated with frequencies $\Omega_1 = \omega_2-\omega_1$, $\Omega_2 = \omega_3-\omega_2$, and $\Omega_3 = \omega_3-\omega_1 = \Omega_1+\Omega_2$ to couple the three modes through direct photonic interband transitions~\cite{winn1999interband,FangPRL2012}. 
The dynamic modulation is applied uniformly along $x$ direction with the spatio-temporal profile:
\begin{equation}
    \label{eq:epsilon_waveguide}
    \epsilon(y,t) = \epsilon_r +
    \sum_{i=1}^3 
    \delta_i(y) \cos (\Omega_i t+\phi_i)
\end{equation}
where $\epsilon_r$ is the static relative permittivity, $\delta_i'$s  are the modulation strengths and $\phi_i'$s are the modulation phases. 
The modulation is applied only to the upper $1/3$ of the waveguide in order to get non-zero coupling coefficients between all three modes. 
The modulation in Eq.~\eqref{eq:epsilon_waveguide} can be implemented by a single standing-wave modulator where the index modulation is uniform in space along the propagation direction.

\begin{figure*}[htp!]
    \centering
    \includegraphics[width=\linewidth]{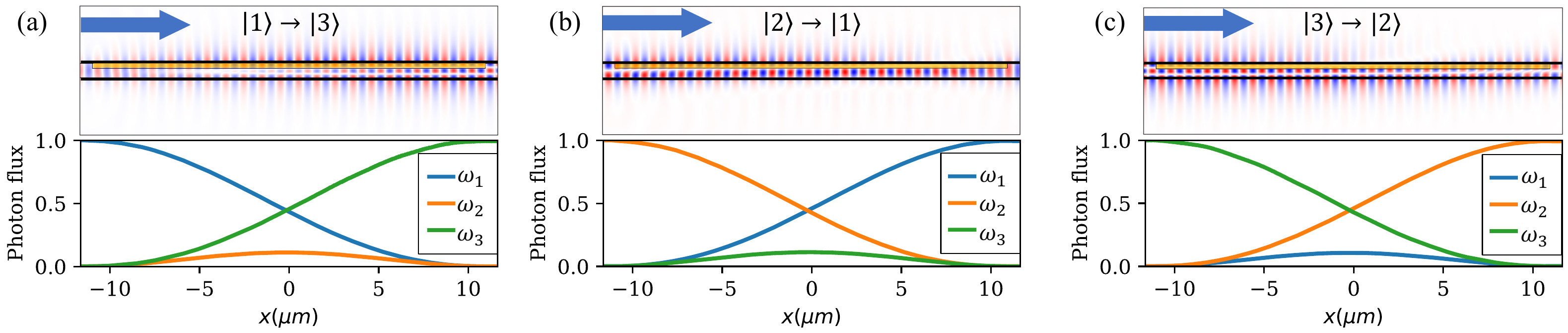}
    \caption{Waveguide simulation results. 
    The top figures are the normalized $E_z$ field distributions with different input  modes  from  the  left port. The solid black lines represent the boundaries of the waveguide and the yellow shaded regions represent the modulation regions.
    The bottom figures are the normalized photon number flux $\abs{a_n(x)}^2$ as a function of the propagation distance $x$. The blue, orange and green lines  represent the photon number flux of modes $\ket{1}, \ket{2}$, and $\ket{3}$, respectively.
    (a), (b), and (c) show the complete conversions from $\ket{1}$ to $\ket{3}$, $\ket{2}$ to $\ket{1}$, and $\ket{3}$ to $\ket{2}$, respectively.
    }
    \label{fig:waveguide_sim}
\end{figure*}

With the above modulation profile, the total electric field inside the waveguide can be written as
\begin{equation}
    \label{eq:wg_Ez_fields}
    E_z(x,y,t) = e^{-j\beta x}
    \sum_{n=1}^3
    a_n(x)\hat{E}_{z,n}(y)e^{j\omega_n t},
\end{equation}
where $n=1,2,3$ represent different modes, $\hat{E}_{z,n}(y)$ is the modal profile of the electric field and $|a_n(x)|^2$ represents the photon number flux~\cite{FangPRL2012,YuNatPho2009} for each mode. 
By substituting Eqs.~\eqref{eq:epsilon_waveguide} and~\eqref{eq:wg_Ez_fields} into the Maxwell's equations, we can derive the coupled mode theory formalism and calculate the transfer matrix of the system~\cite{Sup, haus1984waves, FangPRL2012}.
Suppose 
\begin{equation}
    \label{eq:phase_condition}
    (\phi_1+\phi_2)-\phi_3=\pi/2,
\end{equation}
and 
\begin{equation}
    \label{eq:C_condition}
    C_{12}=C_{23}=C_{13}=\frac{2\pi}{3\sqrt{3}L},
\end{equation}
where $C_{ij} =  \frac{1}{8}\int_{-\infty}^{\infty}dy\hat{E}_{z,j}\delta_i\hat{E}_{z,i}^*$ is the coupling coefficients between different modes~\cite{Sup, FangPRL2012} and $L$ is the modulation length. The transfer matrix of the modulated region then has the form:
\begin{equation}
\label{eq:wg_transfer}
    \mathbf{T} = e^{-j \beta L}
    \begin{pmatrix}
    0 & -1 & 0 \\ 
    0 & 0 & -1 \\ 
    1 & 0 & 0
    \end{pmatrix},
\end{equation}
where $e^{-j \beta L}$ is the global propagation phase.
The transfer matrix indicates strong amplitude non-reciprocity. 
Mode $\ket{1}$ input from the left port is converted to mode $\ket{3}$ at the right port. 
On the other hand, mode $\ket{3}$ input from the right is converted to mode $\ket{2}$, as can be inferred from Eq.~\eqref{eq:wg_transfer} as well as the mirror symmetry of the structure.

Equation~\eqref{eq:wg_transfer} describes a three-mode circulator~\cite{williamson2020integrated}.
In the standard configuration of a circulator, the modes are those of three single-mode waveguides. Here the modes are the three modes of a single waveguide.
Also, in the coupled mode theory we assume only direct transition and ignore indirect transitions that might occur due to the finite length of the modulation region. 
This assumption is validated by the simulation below.

We validate the coupled mode theory analysis above by performing a first-principle multi-frequency frequency domain (MF-FDFD) simulation~\cite{ShiOptica2016}. 
In the simulation, the width of the waveguide is $0.4~\mu$m.
The waveguide has a relative permittivity $\epsilon_r = 12.25$ and is surrounded by air. 
Its dispersion relation for the lowest three modes are shown in Fig.~\ref{fig:waveguide_config_level}(b).
The modulation region has a length of $L = 22.4~\mu$m and a width that is equal to $1/3$ of the waveguide width.
We choose $\delta_1 = 0.060\epsilon_r$, $\delta_2 = 0.044\epsilon_r$, and $\delta_3 = 0.091\epsilon_r$ such that the coupling coefficients $C_{12}=C_{23}=C_{13}$. 
The angular frequencies of the three modes are  $\omega_1=2\pi\times102.9$~THz, $\omega_2=2\pi\times161.2$~THz, and $\omega_3=2\pi\times241.7$~THz. 
The required frequencies of the modulations that drive these transitions can be calculated as  $\Omega_1=2\pi\times 58.3$~THz, $\Omega_2=2\pi\times 80.5$~THz, and $\Omega_3=2\pi\times 138.8$~THz.
The simulation results as shown in Fig.~\ref{fig:waveguide_sim} indicate the amplitude non-reciprocity as predicted from the coupled mode theory formalism. 
Thus we have demonstrated that to construct a device with amplitude non-reciprocity requires only a single standing-wave modulator.
While here for the purpose of reducing computational cost, we have used large modulation strengths and frequencies, these modulators can be designed with realistic modulation strength of $\delta/\epsilon \sim 10^{-3}$, and the modulation frequencies in the $100$~GHz frequency range~\cite{wang2018NatLithium,He2019HighModulator}, using the coupled mode theory formalism.
Under these more realistic assumptions on the modulations, the transfer matrix of the system still has the form of Eq.~\eqref{eq:wg_transfer}, but the device has a longer length of $10$~mm scale. 
By choosing the photonic bands of different waveguide modes to be parallel, the operating bandwidth of the modal circulator can be as broad as on the order of THz~\cite{yu2009integrated}. 
Here for simplicity, we consider lossless system. In the presence of the loss, the coupling constants (i.e. the $C_{ij}$ in Eq.~\eqref{eq:C_condition}) must dominate over the loss rate of the waveguide in order for the circulator to operate.

\begin{figure*}[thpb!]
    \centering
    \includegraphics[width=\linewidth]{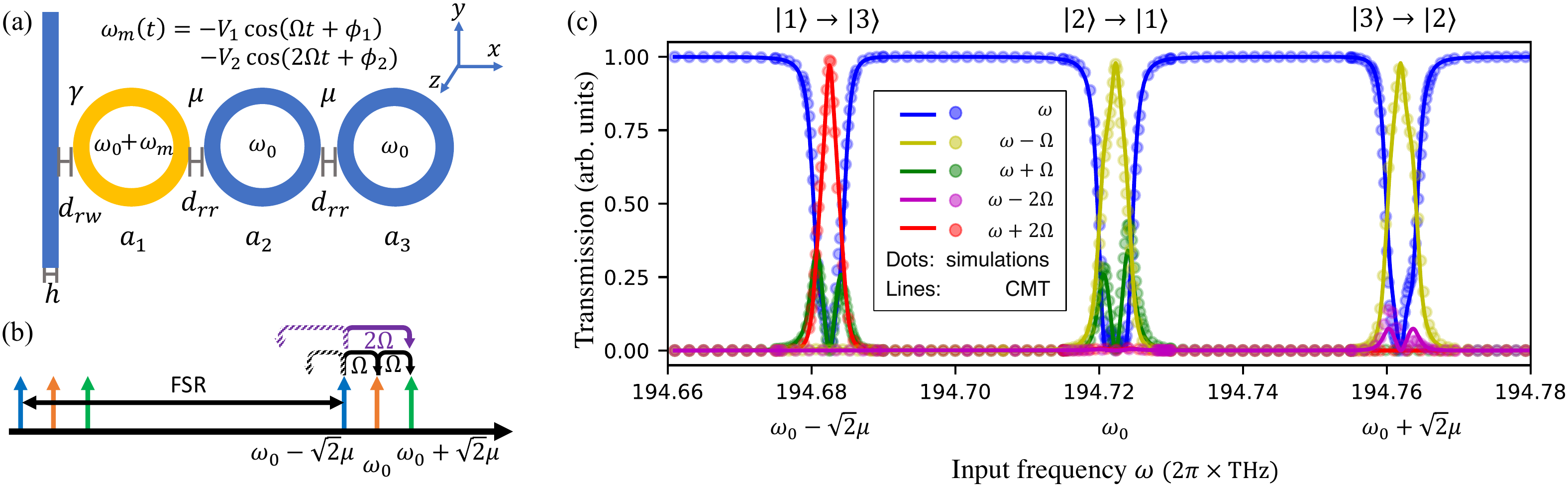}
    \caption{Coupled-ring system. 
    (a) The coupled-ring system preserves mirror symmetry about the $xz$ plane. 
    The system consists of three identical ring resonators coupled with coefficient $\mu$. 
    Only the yellow ring is coupled to a straight waveguide and is modulated such that its resonant frequency varies as $\omega_0+\omega_m(t)$. 
    (b) The frequency spectrum for the coupled-ring system. 
    The coupled-ring system supports three supermodes with frequency separation $\sqrt{2} \mu$, which is far smaller than the free spectral range (FSR).
    (c) Transmission of different frequency components as a function of the input frequency $\omega$. 
    The solid lines are the coupled mode theory (CMT) fitting results and the dots represent the simulation results. 
    When the incident light has frequency $\omega \approx\omega_0-\sqrt{2}\mu\approx 2\pi\times194.682$~THz,
    there is a strong output at $\omega+2\Omega$, indicating the transition $\ket{1}\to\ket{3}$.
    Similarly, the $\ket{2}\to \ket{1}$ occurs at input frequency around $194.722$~THz and the $\ket{3}\to\ket{1}$ occurs at  input frequency around $194.762$~THz.}
    \label{fig:ring_config_cmt_sim}
\end{figure*}

\textit{Modal circulator.} Both the coupled mode theory and the numerical simulation results as shown in Fig.~\ref{fig:waveguide_sim} indicate that the modulated waveguide structure behaves as a modal circulator as described as $\ket{1}\to\ket{3}\to\ket{2}\to\ket{1}$, where each arrow describes an input/output relation.
This modal circulator behavior can be understood by examining the phases associated with various photonic transition processes, as plotted in Fig.~\ref{fig:waveguide_config_level}(c) and (d). 
In order for a photon initially in mode $\ket{1}$ to make a transition to mode $\ket{3}$, there are two possible pathways. 
In the first pathway, the transition can occur through the modulation at frequency $\Omega_3$, with the associated phase factor of $-je^{j\phi_3}$.
Here $\phi_3$ is the modulation phase, and the phase factor $-j$ is a reciprocal phase factor that arises naturally when one computes a scattering matrix $\mathbf{S}$ from a Hermitian Hamiltonian $\mathbf{H}$ through $\mathbf{S} = e^{-j\mathbf{H}t}$~\cite{Sup}.
In the second pathway, the transition occurs in a two-step process, where the mode $\ket{1}$ first makes a transition to mode $\ket{2}$, and then makes a transition to mode $\ket{3}$. 
These transitions are associated with the phase factors $-je^{j\phi_1}$ and $-je^{j\phi_2}$. 
With the choice of the parameters in Eqs.~\eqref{eq:phase_condition} and \eqref{eq:C_condition}, these two pathways constructively interfere, which results in a strong transition from $\ket{1}$ to $\ket{3}$. 
In contrast, in the reversed direction for the photon to make transition from $\ket{3}$ to $\ket{1}$, the reciprocal phase factor of $-j$ remains unchanged, but the phases in the exponents which are associated with the modulation phases change the sign as shown in Fig.~\ref{fig:waveguide_config_level}(d).
Hence the two pathways destructively interfere.
Repeating the process here for other transitions in this three-mode system, we arrive at the modal circulator behavior as indicated in Fig.~\ref{fig:waveguide_config_level}(e).
We note that the reciprocal phase factor of $-j$ plays the role of the reciprocal phase bias~\cite{FangPRL2012,Sup} that allows the non-reciprocity associated with the modulation phase to manifest as amplitude non-reciprocity.

\textit{Coupled-ring system.} 
Based on the discussions above on the mechanisms of modal circulator, we now provide a second implementation using ring resonators, which are more compact as compared to the waveguide design above, but with narrower operating bandwidth.
The system consists of three lossless identical ring resonators with resonant frequency of $\omega_0$, as shown in Fig.~\ref{fig:ring_config_cmt_sim}(a).
The rings are arranged in an array with the same edge-to-edge distance $d_{rr}$, which determines the  coupling coefficient $\mu$ between the rings.
The edge-to-edge distance between the first ring and the waveguide is $d_{rw}$, which controls the decay rate $\gamma$ of the mode in the ring to the waveguide.
Only the first ring (yellow ring in Fig.~\ref{fig:ring_config_cmt_sim}(a)) is coupled to a straight waveguide which provides the input and output ports.

The static system as described above has three supermodes $\ket{1}$, $\ket{2}$, and $\ket{3}$ with resonant frequencies $\omega_1=\omega_0-\sqrt{2}\mu$, $\omega_2=\omega_0$, and $\omega_3=\omega_0+\sqrt{2}\mu$, respectively.
All three supermodes have non-zero field components in the first ring.
Therefore, one can couple all three modes resonantly by modulating only the first ring with the modulation profile:
\begin{equation}
    \epsilon(t) = \epsilon_r + \delta_1 \cos (\Omega t+\phi_1) + \delta_2 \cos (2\Omega t +\phi_2),
\end{equation}
where $\Omega= \sqrt{2} \mu$ describes the fundamental modulation frequency and $\delta_{1,2}$ is the modulation strength in relative permittivity. 
The resonant frequency of the first ring varies accordingly as
\begin{equation}
    \label{eq:ring_omega_t}
    \omega(t) = \omega_0 - V_1 \cos{(\Omega t + \phi_1)} - V_2 \cos{(2\Omega t +\phi_2)},
\end{equation}
where $V_{1,2}$ describes the modulation strength in angular frequency.
Here we assume that $\Omega$ is far smaller than the free spectral range of the ring.

For the incoming field at frequency $\omega\approx\omega_1$ with unit amplitude, we denote the amplitude of the outgoing field at around $\omega+2\Omega\approx\omega_3$ as $t_{31}$, since such outgoing field results from photonic transition in the ring resonators from $\ket{1}$ to $\ket{3}$. 
We can similarly define $t_{13}$ as the transmission from $\ket{3}$ to $\ket{1}$.
Similar to the waveguide system, due to the mirror plane symmetry, the difference between the magnitude of $t_{13}$ and $t_{31}$ indicates amplitude non-reciprocity. 
$t_{13}$ and $t_{31}$ can be calculated analytically with the coupled mode theory~\cite{Sup, Suh2004TCMT, minkov2017exact, peterson2019strongPRL}.
In order to achieve maximum amplitude non-reciprocity, i.e. to have $\abs{t_{31}}=1$ and $\abs{t_{13}}=0$, the modulation must satisfy
\begin{equation}
    \label{eq:ring_cond}
    2\phi_1 -\phi_2 = \pi/2,~~V_1 = V_2 = 2\gamma.
\end{equation}
For which case the photon transition in the ring resonator is unidirectional: the transition from $\ket{1}$ to $\ket{3}$ is allowed whereas the transition from $\ket{3}$ to $\ket{1}$ is forbidden. 
The same condition of Eq.~\eqref{eq:ring_cond} also allows unidirectional photonic transitions for  $\ket{2} \to \ket{1}$ and $\ket{3} \to \ket{2}$.
And thus again, we see that the three supermodes in the ring form a modal circulator similar to the waveguide case.

To verify the analysis above, we perform the MF-FDFD simulations~\cite{ShiOptica2016} and compare the results with the coupled mode theory formalism~\cite{Sup}. 
In our simulation, three identical ring resonators each has $3.875~\mu$m inner radius and $4.125~\mu$m outer radius.
The straight waveguide has a width $h=0.25~\mu$m. 
The ring-ring and ring-waveguide distances are $d_{rr} = 0.32~\mu$m and $d_{rw} = 0.26~\mu$m, respectively.
The whole structure has relative permittivity $\epsilon_r = 12.25$ and is surrounded by air.
The modulation frequency is $\Omega = 2\pi \times 39.62$~GHz to match the supermodes splitting. 
We choose $\phi_1 = \pi/2$, $\phi_2 = \pi/2$, and $\delta_1 / \epsilon_r = \delta_2 / \epsilon_r =  7.14 \times 10^{-5}$ to satisfy the conditions we derived in Eq.~\eqref{eq:ring_cond}.
The parameters for the modulation should be achievable for state-of-the-art electro-optical modulators~\cite{zhang2019molecule,He2019HighModulator,wang2018NatLithium,hu2020reconfigurable, dong2008inducing, chen2014hybrid}. 
The energy cost of these modulators can be as low as 0.1-10 fJ/bit.

We keep five frequency components ($\omega, \omega\pm \Omega,$ and $\omega \pm 2\Omega$) in the MF-FDFD simulations and the coupled mode theory formalism~\cite{Sup}, and plot the normalized steady-state transmissions for different frequency components as a function of the input frequency $\omega$ in Fig.~\ref{fig:ring_config_cmt_sim}(c). 
The agreements between the simulations and the theory indicate that the system indeed operates as a modal circulator within the rings. 
The bandwidth of the coupled-ring system is limited by the decay rate $\gamma$. 
For the above coupled-ring design, the bandwidth is around 10 GHz.
For the lossy ring resonator, the transition rate between various modes must be larger than the total loss rates of the ring resonator in order to maximize the contrast ratio.

\textit{Conclusion.} In conclusion, 
we show that, to achieve amplitude non-reciprocity in dynamically-modulated photonic systems, 
it is necessary to break the spatial inversion symmetry for all previous systems~\cite{FangPRL2012, Lira2012PRL, YuNatPho2009, tzuang2014natphoton} where only two modes are involved in the transport process. 
On the other hand, it is possible to achieve amplitude non-reciprocity with a three-mode system when the spatial inversion symmetry is preserved. 
We numerically demonstrate the concepts using a three-mode waveguide system and a three-coupled-ring system with the support of the coupled mode theory.
Both systems form a modal circulator with only a single standing-wave modulator while preserving the mirror symmetry, which greatly simplify the control and design of on-chip non-reciprocal devices based on dynamic modulations.

\begin{acknowledgments}
The work is supported by a MURI project from the U. S. Air Force of Scientific Research (Grant No. FA9550-18-1-0379).
J. F. H. acknowledges support by the National Science Foundation Graduate Research Fellowship Program (Grant No. DGE-1656518).
The authors would like to thank helpful discussions with Dr. Momchil Minkov, Ms. Zhexin Zhao, Dr. Viktar Asadchy, and Prof. Meir Orenstein. 
\end{acknowledgments}

\bibliographystyle{apsrev4-1}

%

\end{document}